# Electron Transfer between Weakly Coupled Concentric Quantum Rings

I. Filikhin, S. Matinyan, J. Nimmo, and B. Vlahovic

Department of Physics, North Carolina Central University, 1801 Fayetteville Street, Durham, NC 27707, USA

**Abstract:** The electronic structure of the semiconductor double concentric quantum nano-ring (DCQR) is studied under the single sub-band effective mass approach. We show that in the weakly coupled DCQR, that has been placed in transverse magnetic field, the electron spatial transition between the rings can occur due to electron level anti-crossing. The anti-crossing of the levels with different radial quantum numbers provides the conditions when the electron tunneling between rings becomes possible. Results of numerical simulation for the electron transition are presented for DCQRs of different geometry. In particularly, the system of a QR with a QD located at center of this QR is considered.

# 1. Introduction

Quantum Rings (QR) are remarkable meso- and nano-structures due to their non-simply connected topology. Basis of the interest in these structures is ,of course, the famous Aharonov - Bohm (AB) phenomena with their essential influence of vector potential on the quantum mechanical phase of wave function that results in the numerous nontrivial interference effects in the transport properties in an external magnetic field [1].This interest was essentially supported by the progress in the fabrication of the meso and nano structures with a wide range of geometries including single and double rings [2-4] that allows to confront their properties with the atomic ones.

This progress opens effective applications in the different fields including optics, optoelectronics, quantum cryptography, quantum computing, photonics and so on. The interest in the QR roused tremendously in the connection to the problem of the persistent current of the charge and, afterwards, of the spin in mesoscopic objects [5]. The transition from meso -to nano- scale is essential. In the mesoscopic range where the mean free path is much smaller than ring circumference, the scattering still influence the phase coherent transport and macroscopic number of electrons is unavoidable (see, however, in this light important paper [6] where some of these difficulties are avoided). In the nano scale rings the coherence conditions are more favorable and also we may to reduce the problem to the few or even a single electron. This circumstance amplifies the importance of QR in numerous phenomena. This importance is increased when a perpendicular magnetic field is applied to the nano-rings, leading to the novel effects.

Double connectedness of QR plays fundamental role here. Whereas the Quantum Dot (QD) of the corresponding shape (e.g. circular for two dimension (2D), spherical for 3D) has degeneracy in the radial $n$ ( $n$ =0, 1, 2,… ) and orbital $l$ ( $l = \pm 1, \pm 2, \pm 3,...$ ) quantum numbers, QR just due to this double connectedness, in the absence of the magnetic field has only degeneracy in $l$. Application of a perpendicular magnetic field lifts the degeneracy in $l$ also, thus making possible the energy level crossing at the some value of magnetic field $B$ [7]. This fact gives the direct connection between the effective radius of the electron in the ring and magnetic field at the level crossing point, and, as a result, provides the so called jump of electron from one state (lower or higher ) to the another (higher or lower ) with the gain in the energy.

Use of configurations with double concentric QR (DCQR ) leads to a richer picture. One can see the jump of electron (we mean the single electron scheme in the following) -outer or inner - to another - inner or outer ring due to the level crossing at the given $B$. Here we have a close analogy with the atomic physics where the "position " of electron may be changed (with the change of quantum numbers ) under the influence of an electromagnetic field with the emission or absorption of quanta. It is not extra to stress that the level crossing is absent for the QD. Ring geometry ensures the periodic boundary conditions leading to the usual quantization of energy levels. Electron traversing the ring behaves like an electron in the Bloch structure when the variation of potential with the period is similar to the behavior in the circuit around ring.

Magnetic flux threads the inner region of the QR. For the DCQRs the three dimensional treatment is especially important and interesting [8] when one takes into account the rings coupling (e.g. by tunneling). The 3D penetration of the magnetic flux into the rings region results in an aperiodic and saturated AB oscillations in magnetization and also in the dependence on ring sizes and shapes [8,9].

In the light of the above stated, it is not surprise that numerous papers were devoted in recent years to the different aspects of the strain free Double Quantum Rings (DQR) [10-13].

The present work is close in essence to Ref. [13] where the effect of a magnetic field on the energy levels of electron and holes for cylindrical shaped DCQR was determined for fixed size and for radial quantum number $n=1,2$, with orbital quantum number $|l|$ changing from 1 to 4. In the present paper, we visualize interesting features occurring in DCQRs composed of GaAs in an $Al_{0.70}Ga_{0.30}As$ substrate and particularly the single electron transfer from one ring to another under influence of a transverse magnetic field $B$. Therefore, we concentrate here, in contrast with the previous related papers, to the electron transition between the electron levels to different radial quantum number $n$. We will see that in the DCQR that have been placed in a normal magnetic field, the electron spatial transition between the rings can occur due to electron levels anti-crossing. Also we study geometry dependence of energy gap between the anti-crossed levels. The electronic structure of the system represented by a QR with a QD located at center of this QR is also considered.

## 2. Theoretical Considerations

We begin from preliminary remarks. In the study of the physics of QR, as in any meso - and nano - scale objects, the Coulomb interaction between electron and hole is involved, and one needs to disentangle their motion from each other. For relatively small QR, the kinetic motion quantization is much stronger than the one connecting with the Coulomb interaction (rings radii are smaller than the Bohr effective radii of carriers which is of the order of several hundred nm). Due to the very different masses of light electron and heavy hole, they move in different circles independently with different potentials. The small height of the rings reduces the mixing between light electron and heavy hole [13]. Moreover, the electron is more flexible in the inter-ring tunneling, whereas heavy hole is mainly localized in the ring [14]. This gives the real possibility to disentangle the electron and hole motion and consider the Schrödinger equation for the single electron.

Thus, we come to the single electron scheme, and using the single subband approach, which is justified due to the relatively large band gap of GaAs, the problem can be expressed by the following Schrodinger equation

$$\left(\hat{H}_{kp} + V_c(\mathbf{r})\right)\Psi(\mathbf{r}) = E\Psi(\mathbf{r}). \tag{1}$$

Here $\hat{H}_{kp}$ is the single band *kp*-Hamiltonian operator, $m^*(\mathbf{r})$ is the electron effective mass, and $V_c(\mathbf{r})$ is the band gap potential, $V_c(\mathbf{r})=0$ inside the QR and is equal to $E_c$ outside the QR, where $E_c$ is defined by the conduction (or valence) band offset for the bulk. The Ben-Daniel-Duke boundary conditions are used on interface of the material of QR and substrate.

Now if one introduces a constant magnetic field in the z direction ($\mathbf{B} = B\hat{z}$) acting on a particle with a charge $q$ and a vector potential $\mathbf{A} = \frac{1}{2}B\rho\hat{\varphi}$, the Schrödinger equation in cylindrical coordinates becomes:

$$-\frac{\hbar^2}{2}\left(\frac{1}{\rho}\frac{\partial}{\partial\rho}\left(\frac{\rho}{m^*}\frac{\partial\Psi}{\partial\rho}\right) + \frac{1}{m^*\rho^2}\frac{\partial^2\Psi}{\partial\varphi^2}\right) + \frac{i\hbar qB}{2m^*}\frac{\partial\Psi}{\partial\varphi} + \frac{m^*(qB/m^*)^2\rho^2}{8}\Psi -$$

$$-\frac{\hbar^2}{2m^*}\frac{\partial^2 \Psi}{\partial z^2}+V_c(\rho,z)\Psi=E\Psi. \qquad (3)$$

We can separate angle coordinate

$$\Psi_{n,l}(\rho,z,\varphi)=\Phi_{n,l}(\rho,z)e^{il\varphi}, \qquad (4)$$

where $n=1,2,3...$ are radial quantum numbers and $l=\pm 0,\pm 1,\pm 2,...$ are orbital quantum numbers, and have

$$-\frac{\hbar^2}{2}\left(\frac{\partial}{\partial \rho}\left(\frac{1}{m^*}\frac{\partial \Phi_{n,l}}{\partial \rho}\right)+\frac{1}{m^*\rho}\frac{\partial \Phi_{n,l}}{\partial \rho}-\frac{l^2}{m^*\rho^2}\Phi_{n,l}\right)+\frac{\hbar l \omega_c}{2}\Phi_{n,l}+\frac{m^*\omega_c^2\rho^2}{8}\Phi_{n,l}+[V_c(\rho,z)-E]\Phi_{n,l}$$

$$-\frac{\hbar^2}{2m^*}\frac{\partial^2 \Phi_{n,l}}{\partial z^2}=0. \qquad (5)$$

Here $\omega_c=|e|B/m^*c$ is the cyclotron frequency. First magnetic field term in (5) is orbital Zeeman term, the second - so called diamagnetic term. The electron spin Zeeman effect has been ignored here since it is considered to be small (see [10,11], for instance). It leads to the small shift of the crossing points at a given value of $B$.

We used the finite elements method to solve the eigenvalue problem (5) numerically [15].

## 3. Description of the model

In the present paper, as was stated above, double concentric quantum rings (DCQR) composed of GaAs in an $Al_{0.70}Ga_{0.30}As$ substrate under the influence of a magnetic field is studied. For each simulation, the values $m^*=0.067 m_0$ and $0.093 m_0$ were used for the bulk values of the effective masses for the DCQR and substrate respectively. The contribution of strain was ignored in this paper because the lattice mismatch between the rings and the substrate is very small. The confinement potential $V_c(\mathbf{r})$ was considered to be zero in the rings and 0.262 eV in the substrate [3].

To indicate the electron localization in DCQR, we used the electron effective radius $R_{n,l}$ which is defined as root mean square (rms) radius by the relation: $R^2_{n,l}=\int|\Phi^N_{n,l}(\rho,z)|^2 \rho^3 d\rho dz$, where $\Phi^N_{n,l}(\rho,z)$ is the normalized wave function of electron.

The electron in this weakly coupled DCQR can be localized in inner or outer rings. The parameters which define the state of electron in DCQD are the set of quantum numbers $(n,l)$ and parameter of localization $p$ (position) which can have two values "outer" or "inner". Each state can be described by these three parameters $(n,l), p$. We will see that for DCQR the electron spatial transition between the rings can occur due to the electron levels anti-crossing that has been placed in the magnetic field $B$. The anti-crossing, providing a tunneling, is accompanied by changing the quantum numbers $n$ and $p$ of the $(n,l), p$ set.

## 4. DCQR geometry and electron transition between quantum rings

The DCQR is considered to have rotational symmetry about the z-axis with a height of $H$, the widths $D_1$ ($D_2$) for inner (outer) ring;, the rings separation distance $S$, the inner radius $R_1$. Rings have semi-ellipsoidal shapes. An example of the DCQR geometry is presented in Fig. 1. Here $D_1$ keeps at 8 nm while $D_2$ sets to 18 nm, $S$ =5 nm and $H$ =4 nm, $R_1$=5nm. Thus we consider the plane quantum rings with the condition $H \ll D$, what enhances the role of lateral size confinement effect.

Electron transition in the DCQR means the change of electron localization from outer (inner) to the inner (outer) rings. An example of this transition is presented in Fig. 2, where the single electron energies of different quantum states are shown as functions of magnitude $B$ of the magnetic field. The electron transition can occur for three values of $B$: 6 T, 14 T and 18 T. One can assume the initial position (B< 5T) of electron in the (2,-2) state is at outer ring. The localization is changed when the magnetic field reaches the value of 6 T. In this point two energy levels are anti-crossed. The first is with the quantum numbers (2,-2) and the second is (3-2). For $B$ <5 T, the electron of the second state was located in inner ring. At the point of 6 T the electron position change is carried out. If the magnetic field is increasing then the electron takes the energy level with minimal energy. It is the (2-2) level again. However, the position of the electron is the inner ring. The same effect occurred at 14 T (and 18 T) of the magnetic field value when the levels (1, -1) and (2,-1) ((1,-2) and (2,-2)) are anti-crossed. In Fig. 3 we show the changing of rms radius of the electron in the magnetic field for the considered above anti-crossings. It is important to note that the electron position change is not carried out by jump as it can be assumed by Fig. 3a. There is an interval of the magnetic field where the rms radius is smoothly changed from outer ring to inner ring. This is shown in Fig. 3b. One can conclude from Fig. 2 that the electron transition is only possible between levels with different radial quantum numbers.

Another geometry used for DCQR is showed in Fig. 4. This geometry differs from previous one by choosing the equal width of the rings. In Fig. 5 we present the energy curves obtained for single electron in magnetic field with $n = 1,2$, $l = -4,-5,-6$ and $n = 3,4$, $l = -4,-5,-6$. Note that $l = -1$, $n = 3,4$ states belong to the strongly coupled regime and corresponding electron wave functions are spread over whole volume of the DCQ for magnitude of the magnetic field up to $B$ =5 T.

The energy gap attributed to the anti-crossing of the $n,m$ levels is a result of energy splitting of the single level occurred in the system of two coupled quantum objects (see e.g. [16]). Evaluation of this gap $\Delta E_{nm}$ can be performed using the following relation:

$$\Delta E_{nm} \sim \sum_{p=outer,inner} \int \Psi^s_{(n,l),outer}(r,z) V_p(r,z) \Psi^s_{(m,l),inner}(r,z) r dr dz, \qquad (6)$$

Here $\Psi^s_{outer}(r,z)$ is the wave function of the single "outer" QR when the inner QR is absent. $V_p(r,z)$ is the confinement potential in the inner or the outer ring. $\Psi^s_{inner}(r,z)$ is the wave function for "inner" QR without "outer" QR. Here we assume that these wave functions were normalized. In considering quantum rings system, the result of this integration depends on an overlapping wave functions. This overlapping depends on the distance between quantum rings and spreading of the single wave function over the QR geometry region. The last parameter depends mainly on radial and the slightly on orbital quantum numbers. We illustrated these dependences in Fig. 5-6. In Fig. 5 the influence of the radial quantum number on the energy gap is shown. The energy gaps for $n=1,2$ and $n=3,4$ anti-crossings are visibly different. In particularly, the energy gap of $n=3,4$ anti-crossing is larger than for $n=1,2$. This fact can be also observed from Fig. 6a. At the same time, one can see that there is a weak dependence from orbital quantum number as it is shown in Fig. 6a. In Fig. 6b we show dependence of energy gap as a function of the distance of the ring separation. Behavior of this function is obviously an exponential one for small distances which goes to zero for large values of $S$.

## 5 DCQR geometry motivated by ring experimental fabrication

In this section we consider the DCQR geometry motivated by the fabricated DCQR in Ref. [3]. The cross section of the DCQR is presented in Fig. 7a. Structure of the single electron levels with $l \leq 18$ is shown in Fig. 7b for $B=0$. One can see the double sub-bands with $n=1,2$, $n=3,4$ and $n=5,6$. The electron is well localized in outer ring for $n=1$, $l=1,...$, and in inner ring for $n=2$, $l=1,...,3$. The wave functions of the rest states are distributed between inner and outer rings. To illustrate this distribution, the electron energies in DCQR and rms radius $R$ for the states $n=1,...,6, l=0$ are shown in Fig. 8. The electron is strongly localized in outer (inner) ring when $n=1 (n=2)$; 10 nm $< R <$ 30 nm for inner ring and 45 nm $< R <$ 60 nm, for outer. In Fig. 9 the wave functions of weakly and strongly coupled electron levels are shown by the contour plots of the wave functions of the states (1,0) and (2,0), (3,0) and (4,0), (5,0) and (6,0). Strong localization of electron in inner or outer ring has place for the (1,0) and (2,0) states (see also Fig. 3).

Electron states of the sub-bands $n=1,2$ with different orbital momentum $l$ are also well separated by the rings. The electron of the ground state (1,0) is located in inner ring (see Fig. 9). The states with $|l| \neq 0$ are the states of the subband $n=1$ and are lower members of the doublet $n=1,2$. The electron of these states is located in outer ring. We illustrate this fact in Fig. 10 where the excitation energy several low-lying levels is shown along with electron position in DCQR. We can explain the situation, shown in Fig. 10, by competition of two terms of the Hamiltonian of Eq. (3). The first is the term including first derivative of wave function over $\rho$ in differential operator of kinetic energy; the second is the centrifugal term (cf. [10]). For $|l|=0$, when the centrifugal force is absent, the most probably localization of the electron is in inner ring. The centrifugal force is repulsive for electron in inner ring. For $|l| \neq 0$ the electron is localized in outer ring.

Summarizing, one can say that for B=0 the well separated states are only the states $(1,l),p$ and $(2,l),p$. Thus, used notation is proper only for these states. Other states $(n>2,l)$ of this DCQR are strongly coupled states (see Fig. 8-9). Note that in Ref. [13] the $(2,l),p$ and $(1,l),p$ are alternatively denoted as the L and H states, respectively.

Crossing of electron levels in the magnetic field $B$ are presented in Fig. 10. Two crossings can be accompanied by the electron transfer from one ring to another one. The first is at 1 T, another one is at 1.35 T. However at the first value of $B$ there is no electron transfer. Effective radius of electron is not changed. This situation is like when we have crossing levels of two independent rings. There is no tunneling. The radial quantum numbers are the same. Another situation is at 1.35 T. There is the crossing of the states with different $n$, the states (1,-1) and (2,-1) of independent single rings crossed at 1.35 T. In the DCQR this crossing becomes anti-crossing with possibility for electron tunneling between rings. These anti-crossed states are members of the double sub-band which is resulting in double QR system when single QR spectrum is splited [17]. In Fig. 12 we show how the effective radius is changed due to the anti-crossing tunneling. Transformation of the profile of the electron wave function with $B$ is given in Fig. 13.

In the case of planar QRs ($H << R$) the relationship between the energy and the magnetic flux $\Phi$ can approximately be described by the following relation for the ideal quantum ring of radius R in a perpendicular magnetic field B [18]:

$$E_p(l) = \hbar^2/(2m^*R_p^2)(l+\Phi/\Phi_0)^2, \tag{7}$$

where $\Phi = \pi R^2 B$, $\Phi_0 = h/e$, $p$ means inner or outer ring. It is clear that this relation leads to the periodic oscillations of the energy with the Aharonov-Bohm period $\Delta B$, defined by the condition $\Phi/\Phi_0 = -l$ ($\Phi_0 = 4135.7$ Tnm$^2$). The AB-period $\Delta B$ is $\Delta B = \Phi_0/\pi/R^2$. Using rms radius as $R$, one can obtain for the inner ring $\Delta B/2 = 1.73$ T, for the outer ring - $\Delta B/2 = 0.274$ T. The $R$ s are 19.5 nm and 49 nm, respectively. The obtained values for $\Delta B/2$ are corresponding to the level crossing points (see Fig. 11): about 2.2T and 0.25 T, respectively. We obtained rough estimation for $\Delta B/2$ which only qualitatively represents the situation in the Fig. 11. The better agreement we obtained using the radius of most probably localization of electron, defined at the maximum of square of the wave function. The electron mostly localized at 17.8 nm and 48.8 nm for inner and outer rings, respectively. With these values, the single ideal ring estimate leads for $\Delta B/2$ to the values of 2.0 T and 0.275 T.

Other interesting quantum system is the system representing as QR with QD located in the center of the QR. The cross section of the system is given in Fig. 14. In Fig. 15 we present the results of calculations for electron energies of the (1,0) and (3,0) states in the magnetic field $B$. Once more, we can see the situation of the level anti-crossing for $B$ about of 12.5 T. This anti-crossing is accompanied by exchange of electron localization between the QD and the QR. It means that if initial state (for $B$ <12.5 T) of electron was the state (1,0),outer, then the "final" state (for $B$ >12.5 T) will be (1,0),inner. The anti-crossing lifts the degeneracy ensuring tunneling.

We illustrate this transition in Fig. 15 -18. All notations in the figures are the same as in the corresponding previous figures. We restricted the explanation of these figures by the texts in the captions of the figures.

Fig. 15 gives the example of energies dependence on $B$ for the system of Fig. 14. It is quite interesting due to the fact that energy of the dot-localized state grows more slowly than the envelope ring-localized state. It can be figured out (by choosing appropriate geometry) that at the enough large $B$ the dot-localized state becomes the ground state [19]. When the Landau orbit of electron becomes smaller than dot size, the electron can enter the dot without an extra increase of kinetic energy. This trapping of electron in QD may be useful for quantum computing. In Fig. 15 the states (1,0),outer and (3,0),inner are anti-crossed. One can use ideal ring formula (7) to evaluate the value $B_{anti}$ corresponding anti-crossing point, taken into account that the energy of QD (3,0),inner level practically does not depend on $B$ and is equal $E_l^{anti}$ ($E_0^{anti}$ =0.03 eV). If for radius of ideal ring we used rms radius of electron $R$ =25 nm (see Fig. 16 for $B$ < 11 T), then calculation gives the value $B_{anti}$ ~12 T. At the same time, direct calculation in Schrodinger equation gives the value of ~12 T as it is shown in Fig. 15. We have obtained surprising good approximation with the ideal ring formula. For comparison, we evaluated the Aharonov- Bohm period using the Eq. (7) and the ideal ring formula with $R$ =25 nm. The difference between these results is 2.5 T and 2.1 T, respectively. It shows that our 3D planar QR can be well approximated by an ideal ring. Another result of this consideration is that one can control the value $B_{anti}$, as a point of anti-crossing of the levels, by variety of QD geometry.

## Conclusion

In the case of weak coupling between QRs there are two different sub-bands: the first is associated by the inner ring, the second - by the outer ring. One can use these "man-made" bands to describe the electron transition between QRs in DCQR as an inter-band transition. An alternative description was presented in Ref. [13] by means of the introducing of the L and H states. Note that we have this case when we considered the geometry for DCQR presented in Fig. 1-6. Another case considered in present paper is the geometry motivated by the experiments [3] (Fig. 7). Here we have only two states which are well separated, since the other states are mixed by the electron localization (there are no definite positions in either inner or outer rings). This case cannot be considered in the terms of the two bands.

In the present work we make visible main properties of the weakly coupled DCQD established by several level anti-crossings that occurred for the states with different radial quantum number $n$ (double sub-bunds). The picture of the dynamics of the single electron in the DCQR qualitatively is as the follows at $B$ =0, an electron is in the lowest state (lower member of the doublet) with $l = 0$ and maybe several states with $l \neq 0$ are more probable to be found in the inner ring. The numbers of these states depend on the geometry of DCQR. For the higher lying states with $l \neq 0$ the centrifugal potential pushes the electron of the lower member of the doublet towards the external ring.

The situation is changed when a magnetic field is included: diamagnetic term in (5) begins to act. With increasing B, this term equalizes with centrifugal terms and thus ensures the smallest distance between the anti-crossed energy levels. In this situation, the electron can be found with the comparable probability in both rings. After avoiding crossing, the diamagnetic term localizes the electron in the internal ring. As a result, there appears to be oscillations of the energy levels and magnetization.

It is worthy to emphasize that in the present paper, in contrast to the many publications on the subject based on the parabolic confinement potentials, we used more realistic potentials including ones motivated by the experimental fabrication of the QRs [3]. It is known [20,21] that parabolic potentials giving the soft confinement barriers, are failed in the description of some properties of the real systems especially for electron, and are not robust relative to the geometry and mutual location of the rings. The alternative to the parabolic potentials –hard well potentials-- are also far from the real systems and look artificial in many aspects.

Concluding, we say that the fate of the single electron in DCQRs, with a perpendicular $B$, is governed by the structure of the energy levels with their crossing and anti-crossing and changing with magnetic field. The above described behavior is the result of the nontrivial excitation characteristic of the DCQRs.

Effect of the trapping of an electron in QD located at the center of SQR may be interesting from the point of view of quantum computing.

This work is supported by the NSF (HRD-0833184) and NASA (NNX09AV07A).

# The figures

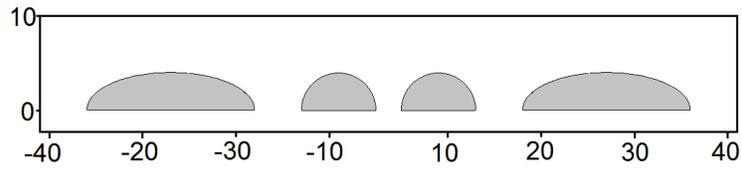

Figure 1. DCQR cross-section. The sizes are given in nm.

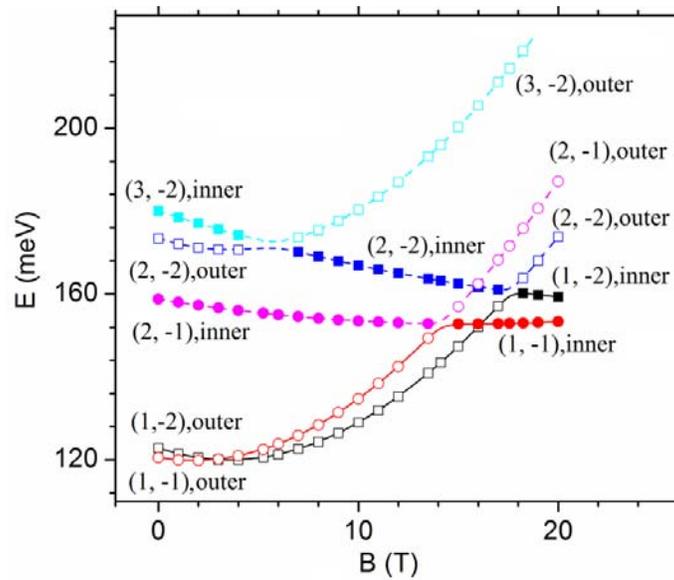

Figure 2. Single electron energies of DCQR as function of magnetic field. Open (solid) symbols mean that the electron is located in outer (inner) ring. Here $D_1$=8 nm, $D_2$=18 nm, $H$ = 4 nm, and $S$ =5 nm (see Fig.1).

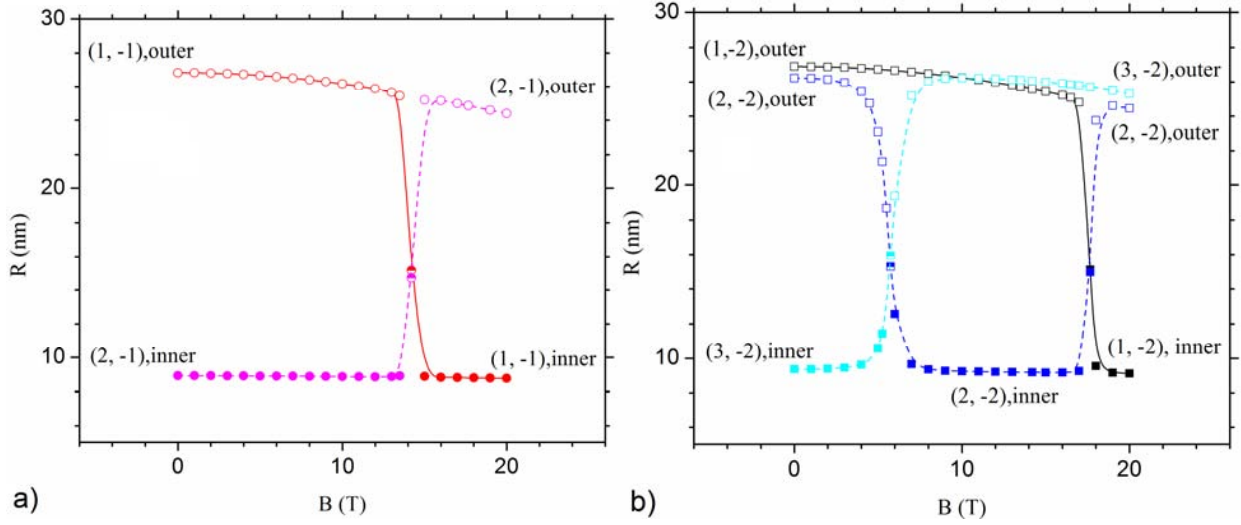

Figure 3. Rms radius of electron in DCQR as function of magnetic field for a) $l=-1$ and b) $l=-2$. Open (solid) symbols mean that the electron is located in outer (inner) ring.

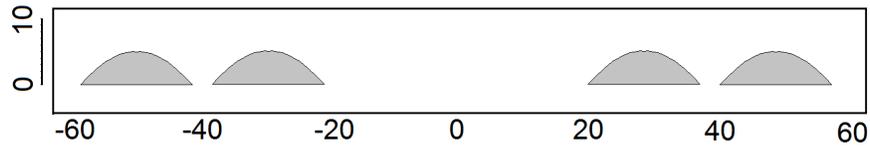

Figure 4. The DCQR cross-section. The sizes are given in nm.

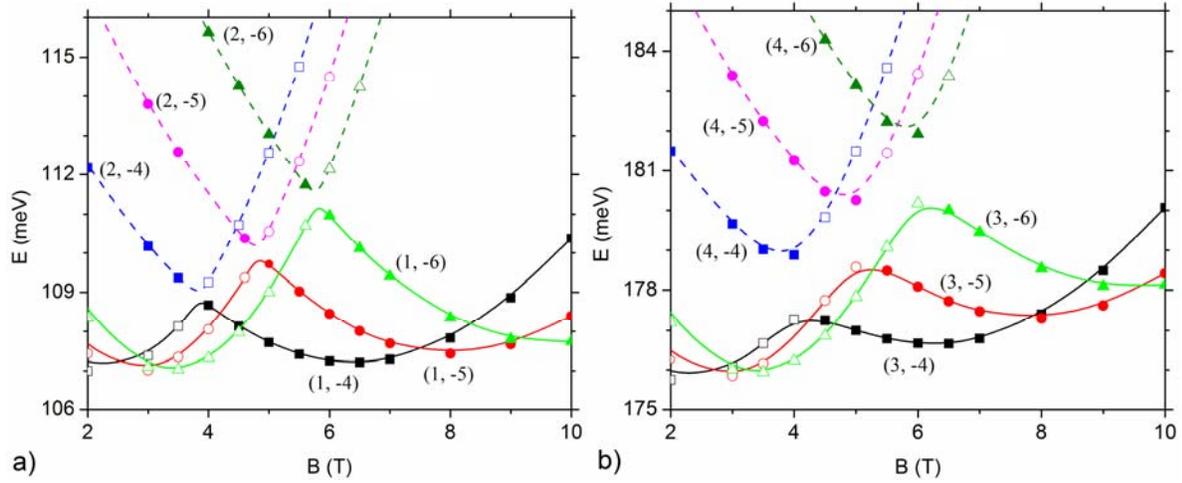

Figure 5. Single electron energies of DCQR as function of magnetic field $B$ for the geometry given in Fig. 4 with $S = 3$ nm. a) The $n = 1,2$ anti-crossing with $l = -4, -5, -6$ and b) the $n = 3,4$ anti-crossing with $l = -4, -5, -6$ are shown. Open (solid) symbols mean that the electron is located in outer (inner) ring.

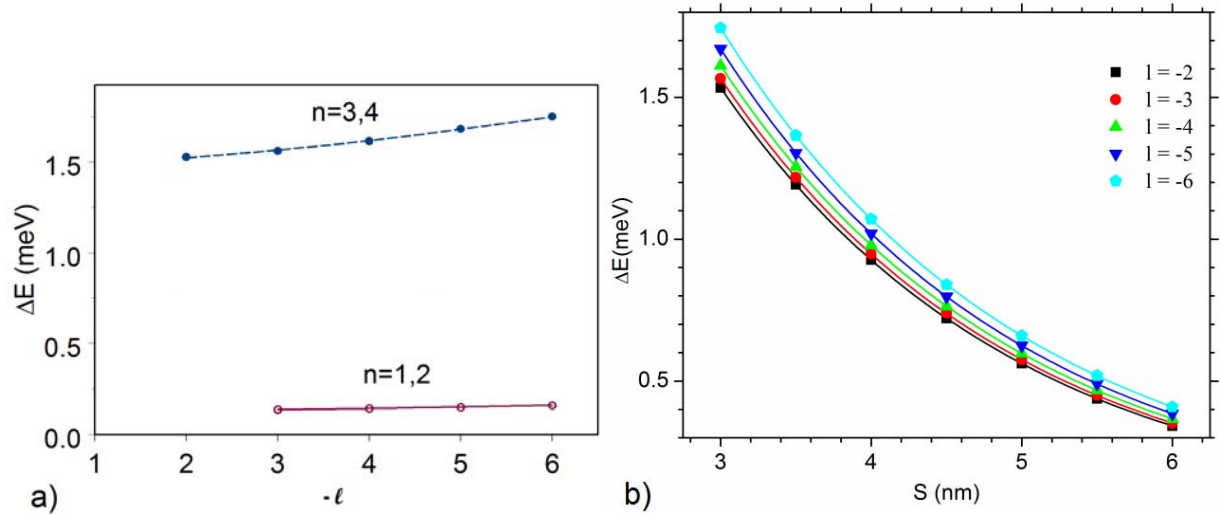

Figure 6. a) Energy gap $\Delta E$ of anti-crossing for $n = 1,2$ (open circles) and $n = 3,4$ (circles) states with different orbital momenta $l$. b) Energy gap $\Delta E$ as a function of ring separation ($S$) for states with $n = 3,4$, $l = -2, -3, -4, -5, -6$.

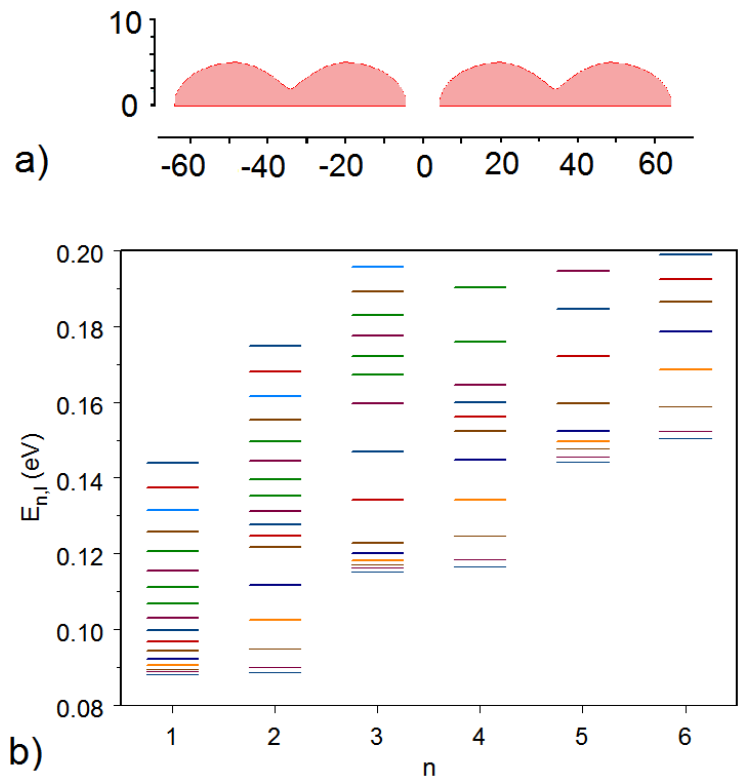

Figure 7. a) Cross section of the DCQR. The sizes are given in nm. b) Single electron energies in DCQR $E_{n,l}$, $n = 1,...,6$, $l = 1,...,18$.

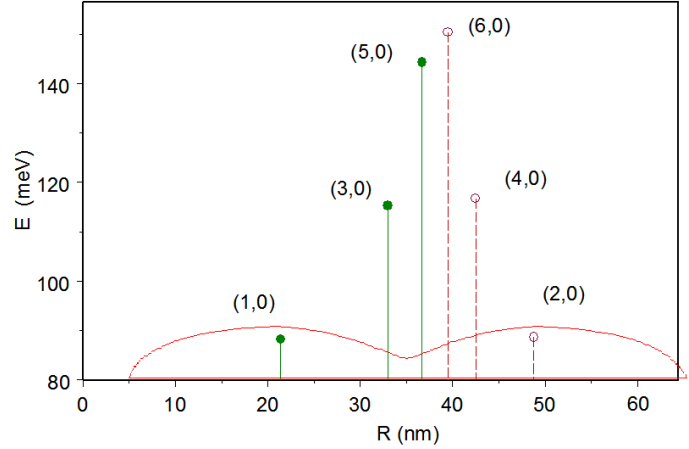

Figure 8. Single electron energies in DCQR and electron rms radius $R$ for the states $n = 1,...,6$, $l = 0$. Contour of the DCQR cross section is shown to note corresponding electron location in DCQR.

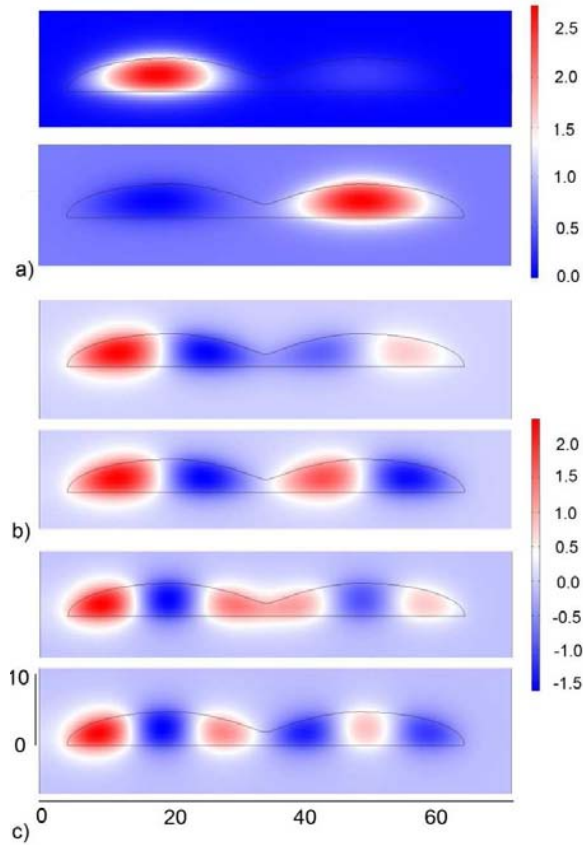

Figure 9. The contour plot of the wave functions of the states are presented: a) (1,0) and (2,0), b) (3,0) and (4,0) c) (5,0) and (6,0). Strong localization of electron in inner (or outer) ring has place for the (1,0) (and (2,0)) states (see Fig. 8). Weakly and strongly coupling regimes in DCQR. Geometry parameters of QRs are chosen to be the same as for experimentally fabricated DCQR

[3]. Right part of cross section of the DCQR is shown (complete picture can be obtained by mirror reflection). The sizes of DCQR along $\rho$ and $z$ directions are given in nm.

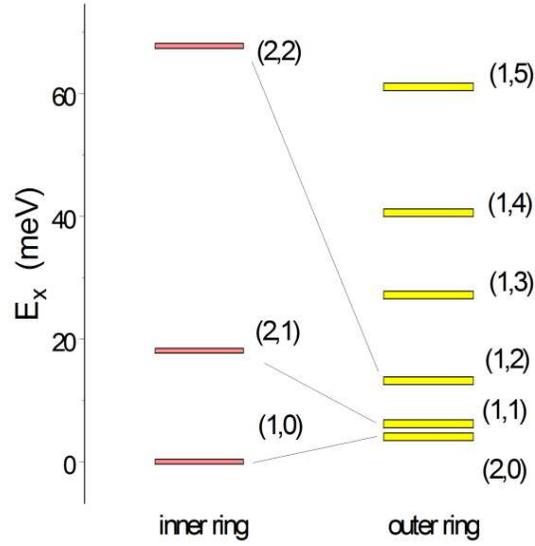

Figure 10. Excitation energy $E_x$ and position of single electron in DCQR for the states $n=1,2$, $l=0,1,2,...5$. The quantum number of each state is shown. Fine lines connect the upper and lower members of the doublets $(1,l)$-$(2,l)$, $l=0,1,2$.

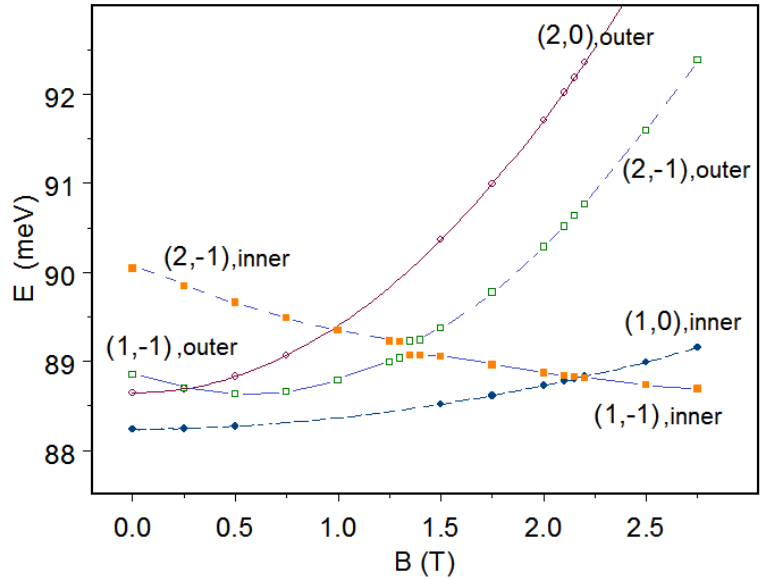

Figure 11. Single electron energies of DCQR as a function of magnetic field magnitude $B$. Open (solid) symbols mean that the electron is located in outer (inner) ring. The quantum numbers of the states are shown.

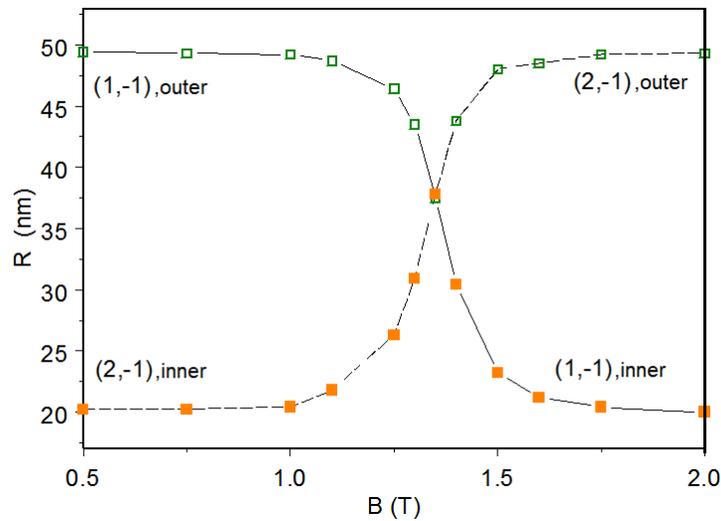

Figure 12. Rms radius of an electron in DCQR as a function of magnetic field for the states $(1,-1)$ and $(2,-1)$. Open (solid) symbols mean that the electron is located in outer (inner) ring. The solid (dashed) line is associated with lower (upper) state of the doublet $n=1,2$.

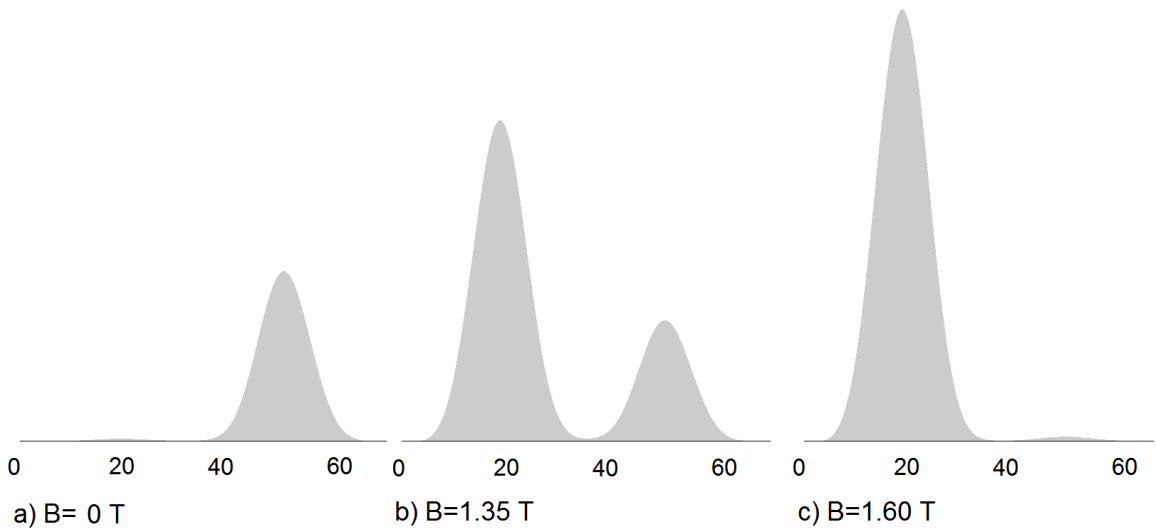

Figure 13. Profiles of the normalized square wave function of electron in the states a) $(1,-1)$,outer; b) $(1,-1)$ or $(2,-1)$ and c) $(1,-1)$,inner for different magnetic field magnitude $B$. a) is the "initial" state ($B=0$) with $R=49.5$ nm, b) is the state of electron transfer ($B=1.35$ T) with $R=37.5$ nm, c) is the "final" state ($B=1.6$ T) with $R=22.0$ nm. The radial coordinate $\rho$ is shown in nm (see Fig. 6 for the DCQR cross section).

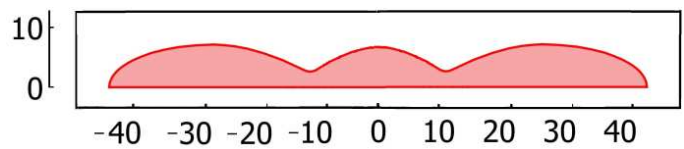

Fig. 14. Cross section of the QR with QD system. Sizes are given in nm.

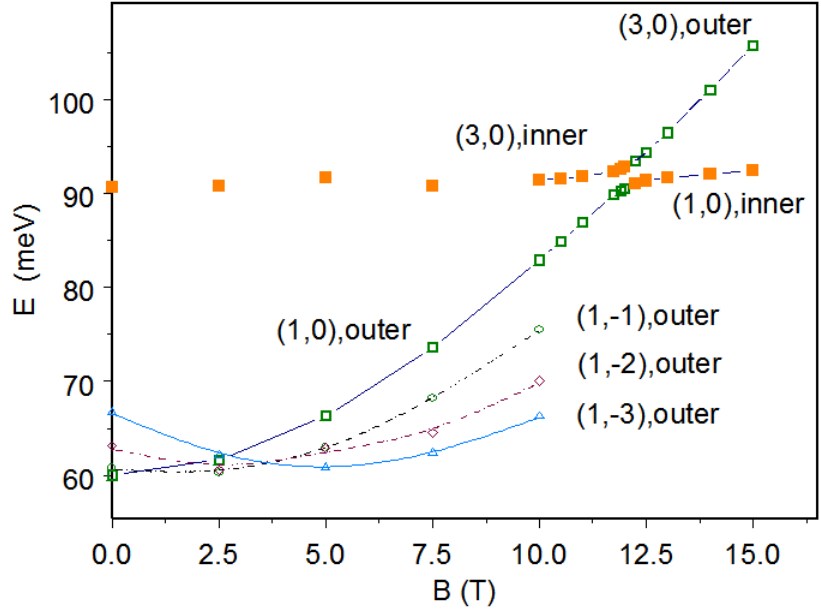

Fig. 15. Energies of the (1,0) and (3,0) states in the magnetic field $B$ for the QR with QD system. The open symbols mean that the electron is localized in the ring. The solid circles mean that the electron localized in QD.

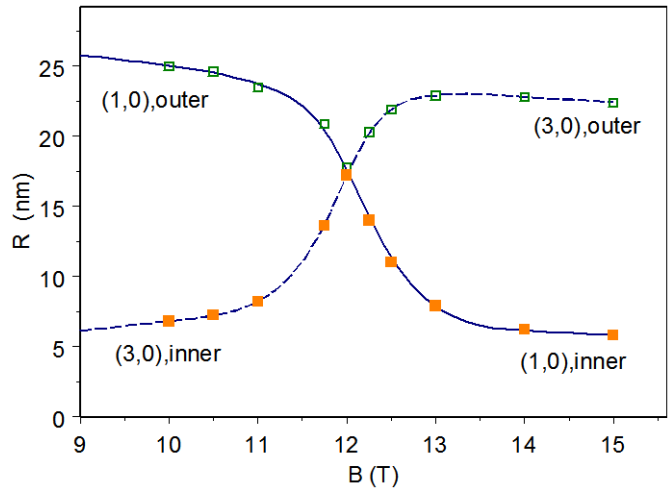

Fig. 16. Rms radius of the single electron states with (1,0),p and (3,0),p in the magnetic field. The solid (dashed) line is associated with lower (upper) state of the doublet.

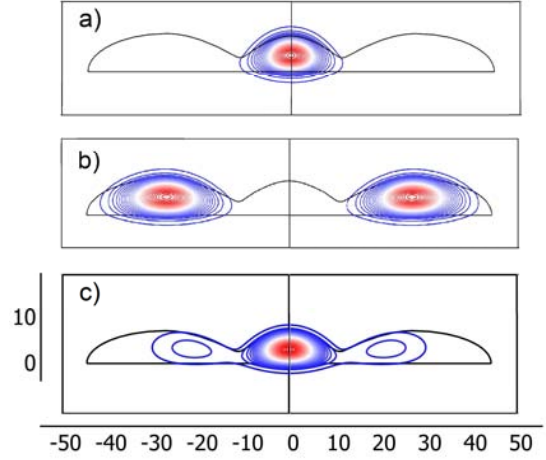

Fig. 17. Contour plots for square wave functions of electron in QD+QR structure. a) of the (3,0),inner state with $B=15$ T(final state), b) of the (1,0),outer state with $B=0$ (initial state), c) of the (1,0) state with $B=12$ T (state with tunneling).

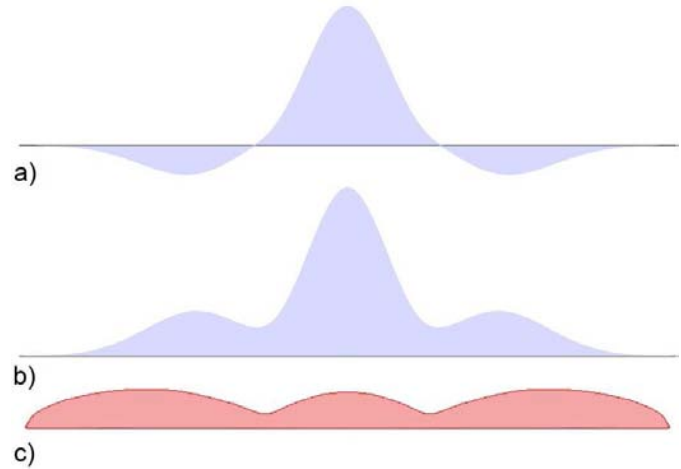

Figure 18. Profiles of wave functions of (3,0) (a) and (1,0) (b) states with $B=12$ T at point of anti-crossing. c) Cross section of the QD+QR structure.